%% file: main.tex
\documentclass[conference]{IEEEtran}
\usepackage{cite}
\usepackage{amsmath,amssymb,amsfonts}
\usepackage{graphicx}
\usepackage{changes}
\definecolor{mygreen}{rgb}{0.0, 0.5, 0.0}  
\setaddedmarkup{\textcolor{mygreen}{#1}}
\definecolor{myred}{rgb}{0.8, 0.2, 0.2}
\setdeletedmarkup{\textcolor{myred}{\sout{#1}}}
\usepackage{textcomp}
\def\BibTeX{{\rm B\kern-.05em{\sc i\kern-.025em b}\kern-.08em
    T\kern-.1667em\lower.7ex\hbox{E}\kern-.125emX}}

\IEEEoverridecommandlockouts
\usepackage{cite}
\usepackage{tabularx}
\usepackage{mathrsfs, amsmath,amssymb,amsfonts}
\usepackage{graphicx}
\usepackage{textcomp}
\PassOptionsToPackage{table,dvipsnames,svgnames,x11names}{xcolor}
\usepackage{xcolor}
\usepackage{colortbl}

\def\BibTeX{{\rm B\kern-.05em{\sc i\kern-.025em b}\kern-.08em
    T\kern-.1667em\lower.7ex\hbox{E}\kern-.125emX}}

\usepackage{lipsum}

\usepackage[utf8]{inputenc}
\usepackage[T1]{fontenc}

\usepackage[abbreviations]{foreign}
\usepackage{algorithm}
\usepackage{algpseudocode}

\usepackage{xspace}
\usepackage{siunitx}
\usepackage{listings}
\usepackage{multirow}

\usepackage{subcaption}
\usepackage{textcomp}
\usepackage{booktabs}

\usepackage{url}
\usepackage{pifont}
\usepackage[colorlinks=true,linkcolor=blue,anchorcolor=blue,citecolor=blue,filecolor=black,menucolor=black,runcolor=black,urlcolor=blue]{hyperref}
\usepackage{orcidlink}

\usepackage[inline]{enumitem}
\setlist{noitemsep,topsep=0pt,parsep=0pt,partopsep=0pt}
\usepackage{textgreek}
\usepackage[capitalise]{cleveref}

\usepackage{tikz}
\usetikzlibrary{positioning, arrows.meta}
\usetikzlibrary{matrix}
\usepackage{pgfplots}

\definecolor{lightcolor}{rgb}{0,0.5,1}

\newcommand{\pie}[1]{%
\protect\begin{tikzpicture}
 \protect\draw (0,0) circle (1ex);
 \fill (1ex,0) arc (0:#1:1ex) -- (0,0) -- cycle;
\protect\end{tikzpicture}%
}

\newcommand{\sys}{\textsc{PhishingHook}\xspace}

\usepackage{fontawesome5}
\newboolean{showcomments}
\setboolean{showcomments}{true}
\ifthenelse{\boolean{showcomments}}
{ \newcommand{\mynote}[3]{
		\fbox{\bfseries\sffamily\scriptsize#1}
		{\small$\blacktriangleright$\textsf{\emph{\color{#3}{#2}}}$\blacktriangleleft$}}}
{ \newcommand{\mynote}[3]{}}

\definecolor{darkgreen}{rgb}{0.3,0.5,0.3}
\definecolor{darkblue}{rgb}{0.3,0.3,0.5}
\definecolor{darkred}{rgb}{0.5,0.3,0.3}

\newcounter{numobserv} 
\definecolor{beaublue}{rgb}{0.88, 0.93, 0.93}
\usepackage{tcolorbox}
\tcbuselibrary{skins, breakable}

\usepackage{etoolbox}
\colorlet{shadecolor}{beaublue}
\tcbset{
	colback=beaublue,
	boxrule=0.5pt,
	boxsep=0pt,
	left=1pt,
	right=1pt,
	top=1pt,
	bottom=1pt,
	sharp corners,
	colframe=white,
}

\usepackage{flushend}

\makeatletter
\let\origsection\section
\renewcommand\section{\@ifstar{\starsection}{\nostarsection}}
\newcommand\nostarsection[1]{\sectionprelude\origsection{#1}\sectionpostlude}
\newcommand\starsection[1]{\sectionprelude\origsection*{#1}\sectionpostlude}
\newcommand\sectionprelude{\vspace{0pt}}
\newcommand\sectionpostlude{\vspace{0pt}}
\let\origsubsection\subsection
\renewcommand\subsection{\@ifstar{\starsubsection}{\nostarsubsection}}
\newcommand\nostarsubsection[1]{\subsectionprelude\origsubsection{#1}\subsectionpostlude}
\newcommand\starsubsection[1]{\subsectionprelude\origsubsection*{#1}\subsectionpostlude}
\newcommand\subsectionprelude{\vspace{-2pt}}
\newcommand\subsectionpostlude{\vspace{-2pt}}
\g@addto@macro\normalsize{%
  \setlength\abovedisplayskip{2pt}
  \setlength\belowdisplayskip{2pt}
  \setlength\abovedisplayshortskip{2pt}
  \setlength\belowdisplayshortskip{2pt}
  \setlength{\floatsep}{3pt}
  \setlength{\textfloatsep}{3pt}
  \setlength{\intextsep}{2pt}
  \setlength{\dblfloatsep}{3pt}
  \setlength{\dbltextfloatsep}{3pt}
}
\makeatother
\usepackage[export]{adjustbox}
\usepackage{eso-pic}

\begin{document}

\title{PhishingHook: Catching Phishing Ethereum Smart Contracts leveraging EVM Opcodes}

\author{
    \IEEEauthorblockN{
        Pasquale De Rosa~\orcidlink{0000-0001-9726-7075}\IEEEauthorrefmark{2},
        Simon Queyrut~\orcidlink{0000-0002-1354-9604}\IEEEauthorrefmark{2},
        Yérom-David Bromberg~\orcidlink{0000-0002-3812-3546}\IEEEauthorrefmark{1},
        Pascal Felber~\orcidlink{0000-0003-1574-6721}\IEEEauthorrefmark{2},        
        and Valerio Schiavoni~\orcidlink{0000-0003-1493-6603}\IEEEauthorrefmark{2}        
    }
    \IEEEauthorblockA{\IEEEauthorrefmark{1}University of Rennes, CNRS, INRIA, IRISA, Rennes, France, \url{david.bromberg@irisa.fr}}
    \IEEEauthorblockA{\IEEEauthorrefmark{2}University of Neuch\^atel, Neuch\^atel, Switzerland, \url{first.last@unine.ch}}
}



\maketitle

\input{authors-version}
\input{abstract}
\begin{IEEEkeywords}
EVM, opcodes, smart contracts, phishing, detection
\end{IEEEkeywords}

\input{intro}

\input{background}

\input{system}

\input{evaluation}
\input{discussion}

\input{rw}
\input{conclusion}

\bibliographystyle{plain}
\bibliography{biblio}

\end{document}

%% file: authors-version.tex

\def\confname{55th Annual IEEE/IFIP International Conference on Dependable Systems and Networks (DSN '25)}
\def\confyear{2025}
\def\confdoi{XXX}

\definecolor{yellowPaper}{HTML}{fff8ae}
\AddToShipoutPictureFG*{%
  \AtTextUpperLeft{%
    \adjustbox{raise=3pt}{
    \begin{tcolorbox}[width=1\textwidth,colback=yellowPaper,enhanced,frame hidden,sharp corners]  
        \centering\scriptsize
        \copyright~\confyear\ 	
         by the Institute of Electrical and Electronics Engineers (IEEE). Personal use of this material is permitted. Permission from IEEE must be obtained for all other uses, in any current or future media, including reprinting/republishing this material for advertising or promotional purposes, creating new collective works, for resale or redistribution to servers or lists, or
         reuse of any copyrighted component of this work in other works.
        This is the author's version of the work.
        The final authenticated version is available online at \href{https://doi.org/10.1109/DSN64029.2025.00033}{https://doi.org/10.1109/DSN64029.2025.00033} 
        and has been published in the proceedings of the 
        \confname.
     \end{tcolorbox}} 
  }%
}%

\hypersetup{
    pdftitle={\copyright~\confyear\  Copyright 2025 by the Association for Computing Machinery, Inc. (ACM). Permission to make digital or hard copies of portions of this work for personal or classroom use is granted without fee provided that the copies are not made or distributed for profit or commercial advantage and that copies bear this notice and the full citation on the first page in print or the first screen in digital media. Copyrights for components of this work owned by others than ACM must be honored. Abstracting with credit is permitted.
    	This is the author's version of the work.
    	The final authenticated version is available online at \href{https://doi.org/10.1145/3583678.3596888}{https://doi.org/10.1145/3583678.3596888} 
    	and has been published in the proceedings of the 
    	\confname.}
}

%% file: abstract.tex
\begin{abstract}
The Ethereum Virtual Machine (EVM) is a decentralized computing engine.
It enables the Ethereum blockchain to execute smart contracts and decentralized applications (dApps). 
The increasing adoption of Ethereum sparked the rise of phishing activities. 
Phishing attacks often target users through deceptive means, \eg fake websites, wallet scams, or malicious smart contracts, aiming to steal sensitive information or funds. 
A timely detection of phishing activities in the EVM is therefore crucial to preserve the user trust and network integrity. 
Some state-of-the art approaches to phishing detection in smart contracts rely on the online analysis of transactions and their traces. 
However, replaying transactions often exposes sensitive user data and interactions, with several security concerns. 
In this work, we present \sys, a framework that applies machine learning techniques to detect phishing activities in smart contracts by directly analyzing the contract's bytecode and its constituent opcodes. 
We evaluate the efficacy of such techniques in identifying malicious patterns, suspicious function calls, or anomalous behaviors within the contract's code itself before it is deployed or interacted with.
We experimentally compare 16 techniques, belonging to four main categories (Histogram Similarity Classifiers, Vision Models, Language Models and Vulnerability Detection Models), using 7,000 real-world malware smart contracts.
Our results demonstrate the efficiency of \sys in performing phishing classification systems, with about 90\% average accuracy among all the models.
We support experimental reproducibility, and we release our code and datasets to the research community.

\end{abstract}

%% file: intro.tex
\section{Introduction}
\label{sec:intro}
Blockchains are distributed ledgers used across several application domains~\cite{chen2018survey, bao2020survey,deepa2022survey}.
They offer well-desired properties, such as anti-tampering, scalability, \etc.
The Ethereum blockchain~\cite{wood2014ethereum} moves nowadays large financial assets and supports an increasing number of financial transactions.
With a market cap of several billion USD~\cite{etherscan:supply}, Ethereum is the 2$^{\texttt{nd}}$ most used blockchain.
Ethereum natively supports sophisticated decentralized applications (\ie \emph{dApps}) implemented by means of special blockchain programs, \ie smart contracts.
The execution of smart contracts operates via the Ethereum Virtual Machine (EVM), a formally-verifiable~\cite{park2018formal} decentralized computing engine.
In a nutshell, the EVM is a stack-based machine that executes \emph{opcodes}, translated to primitive low-level operations such as arithmetic, memory, or stack manipulations (see \S\ref{sec:background}).

The Ethereum blockchain is a frequent target of cybercriminal attacks, extensively studied in the literature~\cite{atzei2017survey,chen2020survey,perez2021smart,10.1145/3643895}. 
Several categories of attack exist: reentrancy attacks (\eg the famous DAO attack~\cite{daoattack} or \cite{grimfinance}, leading to dozens of lost millions), critical reentrancy attacks, default visibility weaknesses, arithmetic under/over flows due to the internal representation of integer values in the EVM, use of custom and malformed random functions~\cite{qian2023demystifying}, front running attacks~\cite{varun2022mitigating}, recent typosquatting attacks~\cite{typosquat}, and more. 
Among these, phishing attacks stand out as one of the most prevalent threats in malicious smart contracts and the main concern for the Ethereum community, as reported by ChainAbuse~\cite{chainabuse}. 
While website phishing attacks have long been a concern~\cite{liu2001introduction,hong2012state}, their rapid growth on popular blockchains has recently attracted significant attention from researchers and practitioners.

We focus on a specific class of phishing attacks, where early detection is crucial to mitigate (or ideally prevent) damage. 
While commercial services for detecting malicious smart contracts are readily accessible to users~\cite{cyvers}, they are often costly.
Although various detection techniques exist (see later in \S\ref{sec:rw}), there is a lack of open-source and reproducible experimental comparisons of their accuracy.


To address this gap, we present \sys, the first framework designed to easily evaluate and compare various techniques for detecting phishing smart contracts, the most predominant type of malicious smart contracts. 
Specifically, we evaluate 16 machine learning (ML) models, including seven -- ViT+R2D2, ViT+Freq, ESCORT, GPT-2 and T5 (in two variants each) -- that have not been previously tested in the context of opcode-based phishing detection for smart contracts. 
The remaining nine models, while having been assessed on related datasets (such as fraud detection containing phishing samples), are evaluated for the first time on a dedicated phishing dataset.

To the best of our knowledge, \sys is the only framework focused exclusively on phishing detection on smart contracts using opcode analysis.
Its architecture, detailed in \S\ref{sec:sys} includes components to extract the bytecode of a smart contract, a bytecode disassembler, a model evaluation module, and finally a post hoc analysis. 
Additionally, \sys provides tools to build or extend existing datasets by crawling public Ethereum explorers or using query services.
The contributions of this paper are:
\begin{itemize}[leftmargin=*]
    \item The construction and public release of the largest dataset of phishing smart contracts targetting the Ethereum blockchain, available at~\cite{DBLP:software/de-rosa2025};
    \item An architecture, prototype implementation and experimental evaluation of \sys;
    \item The analysis of the accuracy of 16 different detection techniques, including statistical approaches, machine-learning methods, computer vision approaches, LLM and vulnerability detectors.
    \item The full set of instructions to reproduce our experiments.
\end{itemize}

\textbf{Roadmap.} 
We overview Ethereum and the required EVM internals in \S\ref{sec:background}.
The architecture of \sys is described in \S\ref{sec:sys}.
We report on our experimental evaluation in \S\ref{sec:eval}.
We detail the lessons learned in \S\ref{sec:lessons}.
We survey related work in \S\ref{sec:rw}, before concluding in \S\ref{sec:conclusion}.

%% file: background.tex

\setlength{\tabcolsep}{3pt}
\begin{table}[!t]
\caption{EVM opcodes for the Shanghai fork (from~\cite{evmcodes}).}
\begin{tabular}{lcll}
\hline
\rowcolor[HTML]{C0C0C0} 
\textbf{Opcode}      & \multicolumn{1}{c}{\textbf{Name}}      & \multicolumn{1}{c}{\textbf{Gas}} & \multicolumn{1}{c}{\textbf{Description}}                      \\ 
\hline
0x00        & {STOP}               & {0}            & {Halts execution}                           \\ 
\rowcolor[HTML]{E3E3E3} 
0x01        & ADD                & 3            & Addition operation                        \\ 
0x02        & MUL                & 5            & Multiplication operation                  \\ 
\textbf{...}         & ...                & ...          & ...                                       \\ 
0xFD        & REVERT             & 0            & Halt execution reverting state changes    \\ 
\rowcolor[HTML]{E3E3E3} 
0xFE        & INVALID            & NaN          & Designated invalid instruction            \\ 
\multirow{2}{*}{0xFF}        & \multirow{2}{*}{SELFDESTRUCT}       & \multirow{2}{*}{5000}  & Halt execution and register \\	      &              &            & account for later deletion \\ \hline
\end{tabular}
\label{tab:opcodes}
\end{table}

\section{Background}\label{sec:background}
\textbf{Ethereum and smart contracts.} Ethereum~\cite{wood2014ethereum} is a state machine where transactions trigger valid state transitions. These transactions are bundled into blocks, cryptographically linked to form a chain. Each block acts as a journal, recording its transactions, the hash of the previous block, and the resulting state. State changes are executed by the Ethereum Virtual Machine (EVM), governed by a \emph{gas} parameter that bounds computation. Transactions come in two forms: \textit{(i)} message calls and \textit{(ii)} contract creation, which deploys accounts with code (\ie smart contracts). Smart contracts are on-chain programs that algorithmically enforce agreements, combining persistent storage with executable functions, typically written in Solidity~\cite{solidity} or Viper~\cite{viper}. Ether (ETH) is Ethereum’s native currency. Consensus on new blocks is achieved through the Beacon Chain using proof-of-stake (PoS). This work focuses on Ethereum starting from the Shanghai update at block 17034870.

\textbf{Ethereum Virtual Machine (EVM).} The EVM~\cite{wood2014ethereum} is a 256-bit stack machine with a maximum of 1024 stack items. Both memory and storage are word-addressed byte arrays, initialized to zero; storage is non-volatile and part of the global state. Execution halts on stack underflows, invalid instructions, or gas exhaustion. Execution is opcode-driven: each opcode represents a specific operation such as arithmetic (ADD, SUB), signed math (SDIV, SMOD), hashing (SHA3), memory/stack manipulation, or contract execution (CALL, DELEGATECALL, etc.). Contracts can be created (CREATE, CREATE2) or removed (SELFDESTRUCT) via dedicated opcodes. As of the Shanghai update, 144 opcodes exist. \autoref{tab:opcodes} lists several, with a complete reference in~\cite{evmcodes}.

\textbf{Phishing attacks in the Ethereum blockchain.}
Phishing attacks in Ethereum involve tricking users into approving harmful transactions or exposing private keys through impersonation of trusted platforms (e.g., dApps~\cite{honeypots}). 
Attackers bait victims with fake incentives (airdrops, staking) via social media or emails, directing them to fraudulent sites mimicking legitimate dApps. 
After wallet connection, victims are prompted to approve a transaction (e.g., "claim reward"), which secretly authorizes attackers to drain their funds.

%% file: system.tex
\section{The \sys Framework}\label{sec:sys}
The architecture of \sys consists of four core modules: \textit{(i)} bytecode extraction module (BEM), \textit{(ii)} bytecode disassembler module (BDM), \textit{(iii)} model evaluation module (MEM), and \textit{(iv)} a post hoc analysis module (PAM). 
An initial data gathering phase leverages etherscan.io~\cite{etherscan} and Google BigQuery \cite{bigquery}. 
We describe the details below.

\begin{figure}[!t]
    \centering
    \includegraphics[scale=0.65]{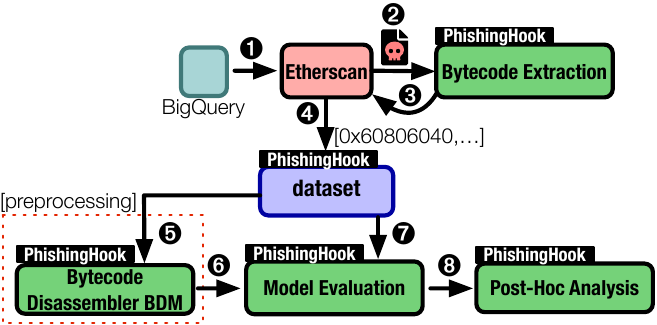}
    \caption{The \sys framework.}
    \label{fig:fw}
\end{figure}

\textbf{Data gathering}. 
We first gather a raw list of unlabeled contract hashes from the Ethereum public dataset available on Google BigQuery (\autoref{fig:fw}-\ding{202}). 
For this study, we limited our search to the contracts deployed between October 2023 and October 2024 ($\approx 4,000,000$). 
As of October 22, 2024, the total number of contracts deployed on the Ethereum blockchain and available on the BigQuery dataset was 68,681,183. 
The public service etherscan.io flags phishing smart contracts with the label ``Phish/Hack'' (\autoref{fig:fw}-\ding{203}). 
We leverage this service to scrape data for each of the 4 million hashes.
Etherscan acts as an independent source of smart contract validation and security analysis. 
Other sources based on community reports (ChainAbuse) are currently proven to be biased~\cite{gomez2024sorting}.

\textbf{Bytecode extraction module (BEM)}. 
The first step of \sys consists in the extraction of the bytecode from the retrieved and labeled contracts. 
To do so, we rely on a public etherscan endpoint (\texttt{eth\_getCode}) \textit{via} an JSON-RPC API (\autoref{fig:fw}-\ding{204}). 
The resulting extracted bytecode constitutes the core dataset adopted in the model training and evaluation phase (\autoref{fig:fw}-\ding{205}).

\begin{figure}[!t]
    \centering
    \includegraphics[width=1\linewidth]{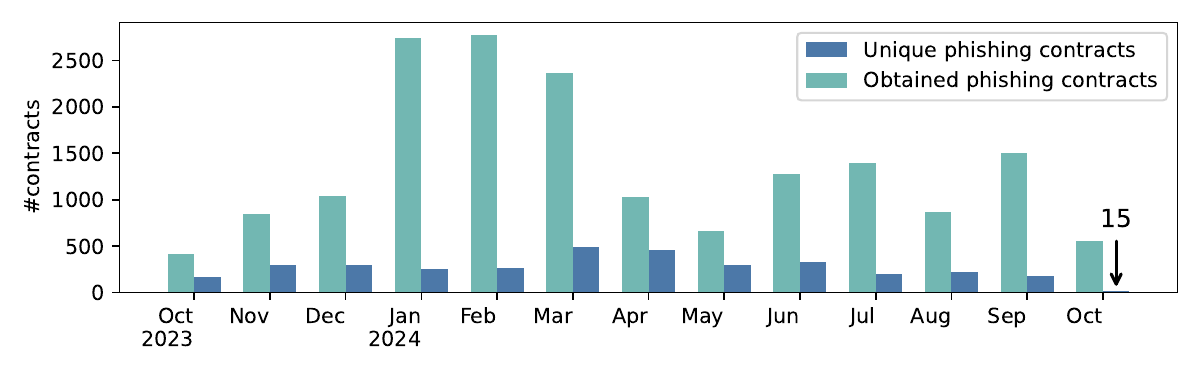}
    \caption{Number of phishing contracts per month over the 2023-10 to 2024-10 period. 
    }
    \label{fig:num_contracts}
\end{figure}

\textbf{Dataset construction}.
From the approximately 4 million contracts collected in the data gathering phase, we sample 17,455 phishing bytecodes, with 3,458 unique bytecodes (\autoref{fig:num_contracts}).
We notice indeed a large majority of duplicate (bit-by-bit) bytecodes.
The reason for such duplicates is the presence of a significant amount of minimal proxy contracts~\cite{proxy}, \ie lightweight and cost-efficient ``clones'' of a main contract, with which they share the same bytecode.
In addition to the malicious smart contracts, we enrich the dataset by a similar number of benign samples (\ie, in our context, not ``flagged'' as malicious on etherscan.io), constituting a final dataset of 7,000 bytecodes. 
We release this novel dataset via our public repository:~\cite{DBLP:software/de-rosa2025}. 
To our knowledge, this is the largest dataset of phishing smart contracts available to the research community.

\begin{figure}[!t]
    \centering
    \includegraphics[width=1\linewidth]{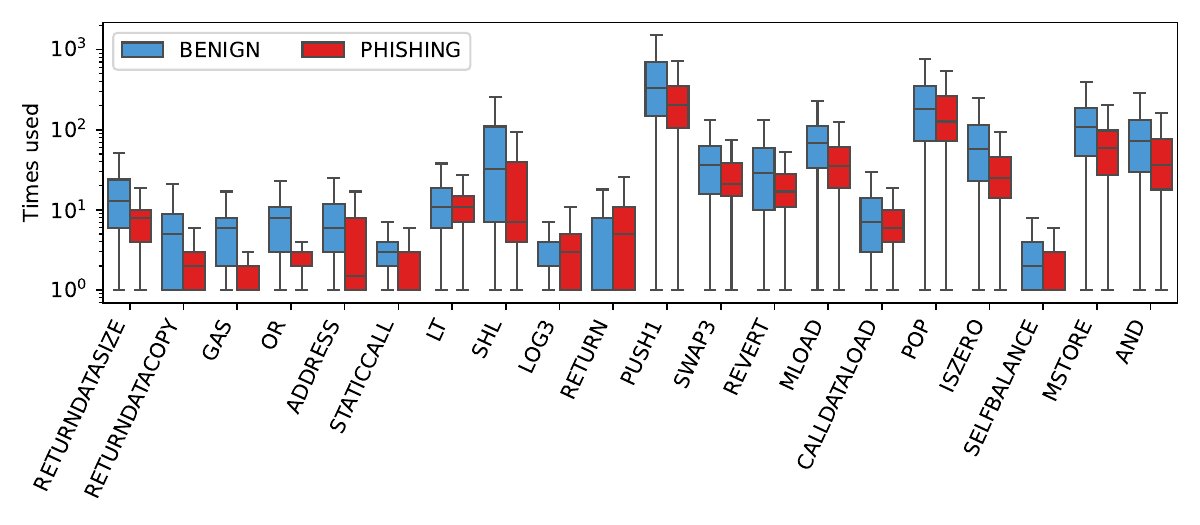}
    \caption{Distribution, by opcode usage, of contracts for 20 opcodes.
    \label{fig:opcode_usage}}
\end{figure}

\textbf{Bytecode disassembler module (BDM)}. 
The deployed bytecode of a smart contract is always publicly accessible on the blockchain, whereas the source code and its AST must be explicitly disclosed by the contract creators. 
Since malicious actors may choose to withhold or obfuscate source code containing harmful functions, we focus on analyzing the deployed bytecode.
The BDM module is in charge of disassembling the bytecode into its corresponding series of opcode instructions (\autoref{fig:fw}-\ding{206}), \ie mnemonic (human-readable alias), operand (argument) and gas (execution cost). 
BDM is a required step only for some of the evaluated detection models (\ie Histogram Similarity Classifiers and ViT + Frequency), that rely on disassembled bytecode features and cannot be trained directly on the original binary. 
For example, a simple bytecode \texttt{0x6080604052} gets disassembled to: \texttt{(PUSH1, 0x80, 3)}, \texttt{(PUSH1, 0x40, 3)}, \texttt{(MSTORE, NaN, 3)}. 
In \sys, the resulting disassembled opcodes are stored in a \texttt{.csv} file for further processing (\autoref{fig:fw}-\ding{207}).  
To perform the disassembling, we leverage a modified version of the Python library evmdasm~\cite{evmdasm}, that we enhanced in order to handle the EVM opcode instructions as of the Shanghai fork. 
The last version of evmdasm registry has been released in March 2022, during the Arrow Glacier update. We added support for two new opcodes: INVALID (that designates an invalid instruction) and PUSH0 (to push 0 bytes in stack). 
We release this enhanced version of evmdasm via our public repository:~\cite{DBLP:software/de-rosa2025}.
One might question whether a discernible pattern exists that could help differentiate phishing contracts from non-phishing ones, based on the relative prevalence of certain opcodes.
We show the distribution of contracts based on how frequently they use each of 20 influential (see \S\ref{sec:eval:shap}) opcodes (\autoref{fig:opcode_usage}).
Phishing contracts utilize opcodes at a similar rate as their benign counterparts, making it unreliable to filter samples solely based on the frequency of a single opcode.

\textbf{Model evaluation module (MEM)}. 
This module comprises the systematic training and evaluation of ML models to classify phishing smart contracts from their bytecode representation (\autoref{fig:fw}-\ding{208}). 
We consider 16 different models from 4 different categories: (i) histogram similarity classifiers (HSCs), (ii) vision models (VMs), (iii) language models (LMs) and (iv) vulnerability detection models (VDMs). 
We compare state-of-the-art, industry-proven models also explored in similar domains, \eg malware detection in smart contracts. 
These include HSCs, ECA+EfficientNet, and SCSGuard, as well as other recent models from different domains, such as ViT, GPT-2, and T5. 
We provide more details on the models and our implementation in  \S\ref{sec:eval:exp} and \S\ref{sec:eval:des}, respectively.
We opted for transformers over traditional sequence models (RNNs) for their superior performance in representation learning for code sequences and detection~\cite{DBLP:conf/emnlp/FengGTDFGS0LJZ20, DBLP:conf/nips/LuGRHSBCDJTLZSZ21}.

\textbf{Post hoc analysis module (PAM)}. 
Finally, we conduct a post hoc analysis on the performances of the ML models (\autoref{fig:fw}-\ding{209}). 
The scope of this analysis is to determine that significant differences exist among the performance metrics of the evaluated models. 
The adoption of the PAM in \sys ensures the statistical validity and significance of the observed results in the evaluation phase. Our scripts for the post hoc analysis are written in R. 
We provide additional details on PAM in \S\ref{sec:eval:pha}.

%% file: evaluation.tex
\section{Evaluation}\label{sec:eval}
\subsection{Experimental setting}
\label{sec:eval:exp}
We use two different GPU-enabled nodes. 
The first features an Intel Core i7-14700KF (28 cores, 5.5 GHz), 64 GiB of RAM, and an NVIDIA GeForce RTX 4090 GPU with 24 GiB of VRAM, utilizing CUDA 12.2. 
The second is powered by an Intel Xeon Platinum 8562Y+ (32 cores, 2.8 GHz), 126 GiB of RAM, and an NVIDIA H100 NVL GPU with 94 GiB of VRAM, running CUDA 12.4.
For deep-learning based models, the design of the train-test pipeline leverages Pytorch v2.5. 
For classical ML models, we use SciKit-learn v1.5. 
The post hoc analysis module uses R v4.4.

\subsection{Compared models}
\label{sec:eval:des}
\sys solves a binary classification task, \ie determining whether the sample is a phishing contract or not.
We next describe the feature extraction methods and models incorporated in \sys. 
As the original authors did not release their implementations as open source, we reimplemented all of them from scratch within the \sys framework.
We include in our experimental study several static code analysis protocols.
We consider 16 methods, detailed next.

\textbf{HSC (Histogram Similarity Classifiers)}~\cite{DBLP:journals/jpscp/LiuPF23}: For each contract bytecode, a histogram of the occurrences of opcodes is created.
It builds a vector of length equal to the number of unique opcodes inside the training set.
The vector is directly served as input (\ie, without normalized nor standardized steps) to several classical ML classifiers proposed by scikit-learn: Random Forest, LightGBM, kNN, XGBoost, CatBoost, Logistic Regression, SVM~\cite{scikit-learn-supervised-learning}.

\textbf{ViT+R2D2 (Vision Model):} Building on methods from Android malware detection research (R2D2~\cite{DBLP:conf/bigdataconf/HuangK18}), we interpret the bytecode as a sequence of hexadecimal color codes.
Each hexadecimal value in the bytecode is mapped to a color in the RGB space.
All pixels (\ie, three channels of integers) are arranged into a $224 \times 224 \times 3$ tensor, with zero-padding applied as needed.
The resulting tensor will serve as input to a classifier.
Due to their widely recognized performance in image classification tasks, we selected the Vision Transformer~\cite{DBLP:conf/iclr/DosovitskiyB0WZ21} to predict the class of the associated contract.
In this study, we utilize a ViT-B/16 model pretrained on ImageNet-1k, with weights obtained from Hugging Face~\cite{huggingface-vit-base-patch16-224} and fine-tuned on our binary classification task.

\textbf{ECA+EfficientNet~\cite{DBLP:journals/cmc/ZhouYWWHL23} (Vision Model):}
The bytecodes of Ethereum smart contracts are transformed into RGB image representations, following a process analogous to ViT+R2D2.
The feature extractor incorporates a modified ECA module~\cite{DBLP:conf/cvpr/WangWZLZH20}, which enhances the model's ability to focus on essential channels.
The ECA mechanism computes a channel-wise attention vector using a depthwise convolution.
The resulting vector is fed to the backbone of the classifier, which leverages a modified EfficientNet-B0 architecture~\cite{DBLP:conf/icml/TanL19}.
Global average pooling is applied to reduce spatial dimensions, followed by a fully connected layer for classification.

\textbf{ViT+Freq (Vision Model):} A lookup table encodes each opcode and operand of the disassembled bytecode to a numerical value which corresponds to their frequency of appearance in the training set.
This frequency is mapped into a color channel intensity value.
The lookup table is constructed exactly once on the entire contract training set.
This frequency encoding performs efficiently as a categorical encoding technique~\cite{DBLP:journals/cstat/PargentPTB22}.
The concept relies on assigning higher pixel intensity values in the R, G, and B channels to the most frequently encountered mnemonics, operands and gas consumptions.
The image tensors have a fixed size of $224 \times 224 \times 3$ which are accordingly zero-padded and subsequently fed to a pretrained ViT-B/16 model (\eg same as in ViT+R2D2).

\textbf{SCSGuard~\cite{DBLP:conf/infocom/HuBX22} (Language Model):} Each hexadecimal string within the bytecode is read as a bigram (sequences of 6 characters). 
These bigrams are numerically encoded to create a vocabulary (\ie, a list of integers), and the sequences are padded to uniform lengths to enable processing by the model.
SCSGuard begins with an embedding layer that maps bigram indices to dense vectors.
A multi-head attention mechanism is applied to capture dependencies between different parts of the sequence, followed by a GRU layer that models sequential patterns in the data.
Finally, a fully connected linear layer generates the logits, representing the model's predictions.

\textbf{GPT-2~\cite{radford2019language} and T5~\cite{DBLP:journals/jmlr/RaffelSRLNMZLL20} (Language Models):}
GPT-2 and T5 are two Transformer-based language models pretrained on a large corpus of text. We selected the latest versions of the models that we could retrieve from the HuggingFace library \cite{hf}.
Despite their modest size (up to 1.5 billion parameters),  these models can be leveraged to build good classifiers~\cite{DBLP:conf/emnlp/NogueiraJPL20, DBLP:journals/corr/abs-2012-13400}. 
Since transformer-based models interpret text as a sequence of tokens, we use the suitable \texttt{GPT2Tokenizer} and \texttt{T5Tokenizer} classes from Hugging Face to process the textual inputs.

\textbf{ESCORT~\cite{escort} (Vulnerability Detection Model):} This system is originally designed to detect code vulnerabilities in smart contracts, and is hereby used for the first time in the context of fraud detection. ESCORT embeds the smart contract bytecode into a vector space. The generated feature representations are then processed by a deep neural network (DNN) model for further analysis. The design supports two operational modes: an initial training phase, where the model is trained to classify individual vulnerabilities in a multi-class labeling context, and a second phase, where the aim is to detect new vulnerabilities through transfer learning.

\subsection{Hyperparameter search}
We rely on Optuna~\cite{DBLP:journals/corr/abs-1907-10902}, an open-source hyperparameter optimization framework designed to automate and streamline the tuning process in ML workflows, to select hyperparameters for all models.
Optuna uses metaheuristics to find the best hyperparameters for models by implementing a define-by-run API, which allows users to dynamically construct search spaces. 
We conducted grid search over an arbitrary search space on the same task as the main evaluation, using 10-fold cross-validation to ensure robust performance assessment. 

\subsection{Results}
\label{sec:eval:res}
To ensure stability of the results, we performed a 10-fold cross-validation over 3 runs, for a total of 30 experiments per model (not including the hyperparameter runs). 
\autoref{tab:metrics} shows the results, averaged over 10 folds and 3 runs. 
We evaluate two variants of GPT-2 and T5: \textalpha{}, where opcode sequences are truncated to fit model token limits and GPU constraints (RTX 4090), and \textbeta{}, trained on an H100 NVL, where full bytecodes are processed in chunks using a sliding window. SCSGuard, relying on n-grams, remains unaffected.
Our results indicate that the vulnerability detector ESCORT is not effective when adapted to a different task like the classification of phishing smart contracts, mostly due to inner limitations of VDMs when applied to social engineering vulnerabilities like phishing, since they exploit human behavior rather than technical flaws in the code.

The most accurate models are the HSCs, with an average Accuracy of 91.52\%, F1 Score of 91.44\%, Precision of 91.61\% and Recall of 91.32\%. 
In particular, the Random Forest approach is both the best performing model overall and the best performing HSC. 
The LMs ranked 2$^{\texttt{nd}}$, with an average Accuracy of 88.83\%, F1 Score of 88.17\%, Precision of 89.50\% and Recall of 88.07\%. SCSGuard is the best performing LM model.
The VMs reported an average Accuracy of 83.75\%, F1 Score of 83.40\%, Precision of 83.26\% and Recall of 83.63\%.
ECA+EfficientNet is the best performing vision model.
All models achieve reasonable performance in detecting phishing from bytecode, with an average Accuracy of 89.07\%, F1 Score of 88.74\%, Precision of 89.24\% and Recall of 88.70\%.

Our results align with the original papers. For HSCs~\cite{DBLP:journals/jpscp/LiuPF23}, Random Forest was also the best model but had slightly lower accuracy (85.17\%). 
In contrast, ECA+EfficientNet~\cite{DBLP:journals/cmc/ZhouYWWHL23} achieved 98.2\%. However, these evaluations were conducted on broader fraud datasets, making direct comparison with our results, focused solely on phishing, challenging. SCSGuard~\cite{DBLP:conf/infocom/HuBX22} detected four out of five phishing scams, with a lower accuracy (80\%). Other models were our own implementations; ESCORT, designed for vulnerability detection, had not been used for fraud detection.
The discussed metrics are crucial for demonstrating the performance of a malware analysis system~\cite{journals/compsec/UcciAB19}.
 However, the trade-offs with cost metrics, such as training/inference time in relation to data size, must also be considered; they're addressed in greater detail in \S\ref{sec:eval:sca}.

\setlength{\tabcolsep}{5pt}
\begin{table}[t!]
\caption{Averaged performance metrics for the models supported in \sys (best values in bold). \text{\dag}: Histogram, \text{\ddag}: Vision, *: Language, \text{\S}: Vulnerability.
\label{tab:metrics}}
\centering
\begin{tabular}{lcccc}
\hline
\rowcolor[HTML]{C0C0C0} 
{\textbf{Model}} & \textbf{Accuracy (\%)} & \textbf{F1 Score} & \textbf{Precision} & \textbf{Recall} \\ \hline
Random Forest \text{\dag}      & \textbf{93.63} & \textbf{93.49} & 94.23          & 92.76          \\ 
\rowcolor[HTML]{E3E3E3} 
k-NN \text{\dag}                & 90.60          & 90.62          & 89.31          & 91.99          \\ 
SVM \text{\dag}               & 92.60          & 92.32          & \textbf{94.53} & 90.21          \\ 
\rowcolor[HTML]{E3E3E3} 
Logistic Regression \text{\dag} & 83.91          & 84.13          & 82.03          & 86.38          \\ 
XGBoost \text{\dag}             & 93.43          & 93.30          & 93.74          & \textbf{92.88} \\ 
\rowcolor[HTML]{E3E3E3} 
LightGBM \text{\dag}            & 93.39          & 93.26          & 93.80          & 92.73          \\ 
CatBoost \text{\dag}            & 93.10          & 92.95          & 93.62          & 92.30          \\ 
\rowcolor[HTML]{E3E3E3} 
ECA+EfficientNet \text{\ddag}   & 86.63          & 86.16          & 86.88          & 85.52          \\ 
ViT+R2D2 \text{\ddag}          & 85.52          & 85.14          & 85.20          & 85.15          \\ 
\rowcolor[HTML]{E3E3E3} 
ViT+Freq \text{\ddag}            & 79.11          & 78.90          & 77.71          & 80.23          \\ 
SCSGuard *           & 90.46          & 90.12          & 90.95          & 89.35          \\ 
\rowcolor[HTML]{E3E3E3} 
GPT-2\textsubscript{\textalpha{}} *                & 89.95          & 89.60          & 90.39          & 88.91          \\ 
T5\textsubscript{\textalpha{}} *                  & 89.67          & 89.28          & 90.25          & 88.35          \\
\rowcolor[HTML]{E3E3E3} 
GPT-2\textsubscript{\textbeta{} *}                &  88.65         &   88.36       &   88.40       &   88.36        \\ 
T5\textsubscript{\textbeta{} *}                  &  85.41         &  83.47       &  87.49        &  85.40         \\
\rowcolor[HTML]{E3E3E3} 
ESCORT \text{\S}    & 55.91 & 55.82 & 55.78 & 55.91 \\ \hline
\end{tabular}
\end{table}

\begin{figure*}
    \centering
    \includegraphics[scale=0.35]{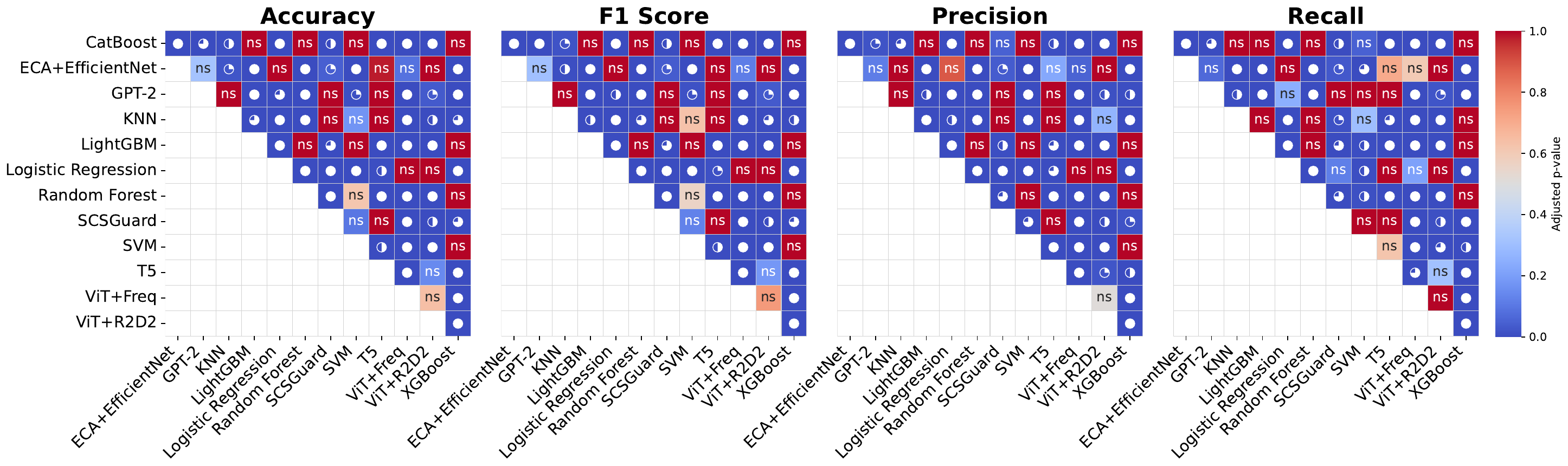}
    \caption{Dunn's test for pairwise comparison between each model's metrics. Significant if $p_{\text{adj}} < 0.05$. Significance levels from \pie{360} (highly significant) down to \pie{90} (mildly significant). Non-significant results are labeled as \textbf{ns}.}
    \label{fig:dunn}
\end{figure*}

\subsection{Post hoc analysis}
\label{sec:eval:pha}
In this section, we describe the methodologies applied for the post hoc analysis of results.
Specifically, our aim is to rigorously compare the performance metrics (Accuracy, Precision, Recall, and F1 Score) obtained for each model to assess the significance of the observed differences.

We excluded from the post hoc analysis the vulnerability detector ESCORT, as it did not perform well on the phishing detection task. 
We also excluded GPT-2\textsubscript{\textbeta{}} and T5\textsubscript{\textbeta{}}, as they were the worst-performing variants of these models.
We conducted the post-hoc analysis on the full experimental results - namely, 30 trials per model (10-fold cross validation $\times$ 3 runs). 
As a result, the number of observation for each metric was 390 (13 models $\times$ 30 trials).
Initially, we tested the normality of each metrics distribution with the Shapiro-Wilk (S-W) test~\cite{shw}.
This represents a crucial step: the following choices of which test to adopt for the group comparison (parametric \emph{vs.} non-parametric) are based on this assumption.
As of the S-W, the test statistic $W$ (a numerical value that indicates the likelihood of observing the results under the null hypothesis) is computed as $\sum_{i=1}^{n} a_i x_{(i)}^2 / \sum_{i=1}^{n} (x_i - \bar{x})^2$, where $x_{(i)}$ are the ordered sample observations, $a_i$ are the coefficients, $\bar{x}$ is the sample mean, and $n$ the sample size. 

The null hypothesis of normality is rejected for values of $W$ significantly lower than 1, such that $p < 0.05$.
In our case, the assumption of normality was violated for 20 model-metric pairs (out of 52).
Hence, we opted for the non-parametric Kruskal-Wallis (K-W) test~\cite{krw}.  
K-W is used to determine whether there are statistically significant differences between the medians of three or more independent groups (in our case, the performance metrics for each of the 13 models). 

The K-W test statistic, conventionally named $H$, is computed as $12/(N(N+1)) \cdot \sum_{i=1}^k R_i^2/n_i - 3(N+1)$, where $k$ are the number of groups, $n_i$ the number of observations for the $i^{\text{th}}$ group, $N$ the total number of observations (for all groups combined) and $R_i$ the sum of ranks, which represent the positions of data points in the $i^{\text{th}}$ group.

\setlength{\tabcolsep}{5pt}
\begin{table}[t!]
\caption{Results of performance metrics with the Kruskal-Wallis test. Significant if $p_{adj} < 0.05$.}
\centering
\footnotesize
\begin{tabularx}{0.7\columnwidth}{Xccc}
\hline
\rowcolor[HTML]{C0C0C0} 
\textbf{Metric} & \textbf{H} & \textbf{p} & \textbf{p\textsubscript{adj}}\\ \hline
Accuracy  & \text{360.81} & $\text{7.35} \times \text{10}^{\text{-70}}$ & $\text{2.94} \times \text{10}^{\text{-69}}$\\ 
\rowcolor[HTML]{E3E3E3} 
F1 Score  & \text{359.78} & $\text{1.21} \times \text{10}^{\text{-69}}$ & $\text{3.63} \times \text{10}^{\text{-69}}$\\ 
Precision & \text{345.21} & $\text{1.44} \times \text{10}^{\text{-66}}$ & $\text{2.88} \times \text{10}^{\text{-66}}$\\ 
\rowcolor[HTML]{E3E3E3} 
Recall    & \text{322.03} & $\text{1.10} \times \text{10}^{\text{-61}}$ & $\text{1.10} \times \text{10}^{\text{-61}}$\\ \hline
\end{tabularx}
\label{tab:kruskal-wallis-results}
\end{table}

The null hypothesis that all the performance metrics have the same median is rejected for values of $H$ significantly high, such that $p < 0.05$.
In this test, we adjust $p$ to $p_{adj}$ with the Holm-Bonferroni correction to reduce the risk of false positives~\cite{aickin}.
As shown in \autoref{tab:kruskal-wallis-results}, the null hypothesis is firmly rejected for all four metrics, thus proving the existence of significant differences between the model performances. 
For this reason, we completed our post hoc analysis performing a Dunn's test~\cite{dunn} with the Holm-Bonferroni correction to determine exactly which model pairs diverge, for any of the four metrics.
This is the appropriate nonparametric pairwise multiple comparison procedure when a Kruskal–Wallis test is rejected~\cite{dunn2}. 
The Dunn's test statistic, conventionally named $Z$, is computed as $(\bar{R}_i -  \bar{R}_j)/(\sqrt{(N(N+1))/12) \cdot (1/n_i + 1/n_j)}
$, where $(\bar{R}_i - \bar{R}_j)$ is the difference in mean ranks between the two group pairs $i$ and $j$, and the denominator is the variance term that accounts for the variability in ranks, adjusted for sample sizes of the pairs $(n_i, n_j)$.

\autoref{fig:dunn} reports these results.
For Accuracy, F1 Score and Precision, the percentage of model pairs that showed significant differences in performance was 65.38\%, while for Recall it was 61.54\%.
For model pairs belonging to the same category (\eg, SVM and k-NN are both HSC while GPT-2 and T5 are both LMs) the percentage of significant differences was lower, \ie 37.04\% for Accuracy and F1 Score, 40.74\% for Precision and 33.33\% for Recall.
Instead, for model pairs belonging to different categories, this percentage increased substantially, \ie, 80.39\% for Accuracy and F1 Score, 78.43\% for Precision and 76.47\% for Recall.
On average, the observed results are extremely significant for model pairs not belonging to the same category, while for similar models there was less statistical relevance in the observed divergencies.

\subsection{Model scalability analysis}
\label{sec:eval:sca}
\begin{figure}[!t]
    \centering    
    \includegraphics[width=\linewidth]{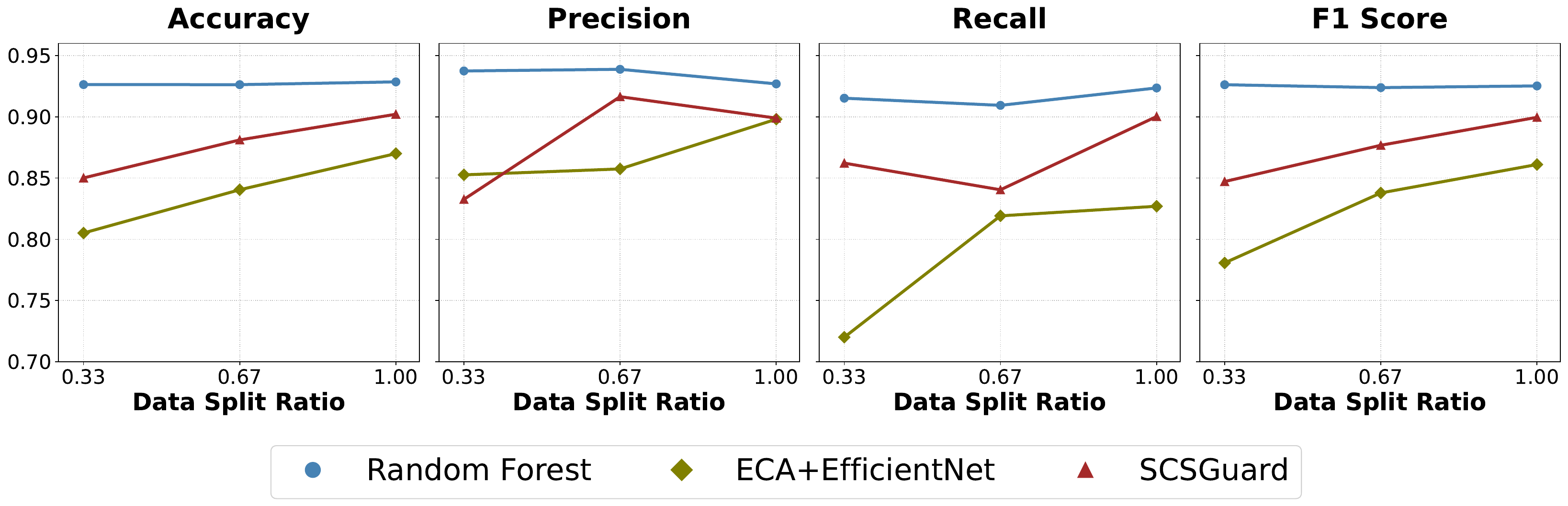}
    \caption{Performance metrics of the best models per data split.}
    \label{fig:scalmetrics}
\end{figure}

Here we conduct a model scalability evaluation.
The goal is to study the impact of data size on the performances of the considered models.
We generated three data splits, containing respectively 1/3, 2/3 and all the smart contract samples.
We trained and evaluated three models (SCSGuard, ECA+EfficientNet, Random Forest), \eg the most accurate models in each category, respectively LM, VM and HSC.
\autoref{fig:scalmetrics} reports our results.
Random Forest is the most accurate model for each of the data splits and its performances remain stable.
SCSGuard and ECA+EfficientNet scale better when increasing the number of training samples. 
This suggests that by increasing the available contract bytecodes in the dataset, Vision and Language models will outperform the HSCs.

\begin{figure}[!t]
    \centering
    \begin{minipage}{0.49\linewidth}
        \centering
        \includegraphics[width=0.84\linewidth]{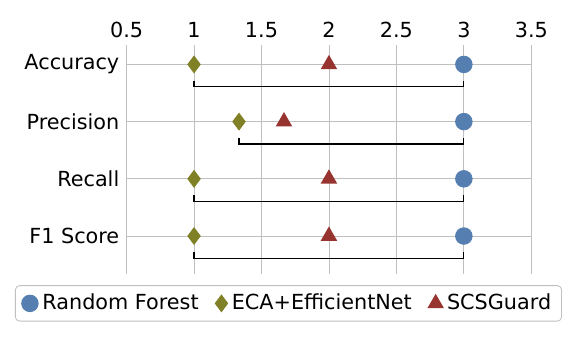}
        \caption{Critical difference diagram of model scalability.}
        \label{fig:cdd}
    \end{minipage}
    \hfill
    \begin{minipage}{0.49\linewidth}
        \centering
        \includegraphics[width=\linewidth]{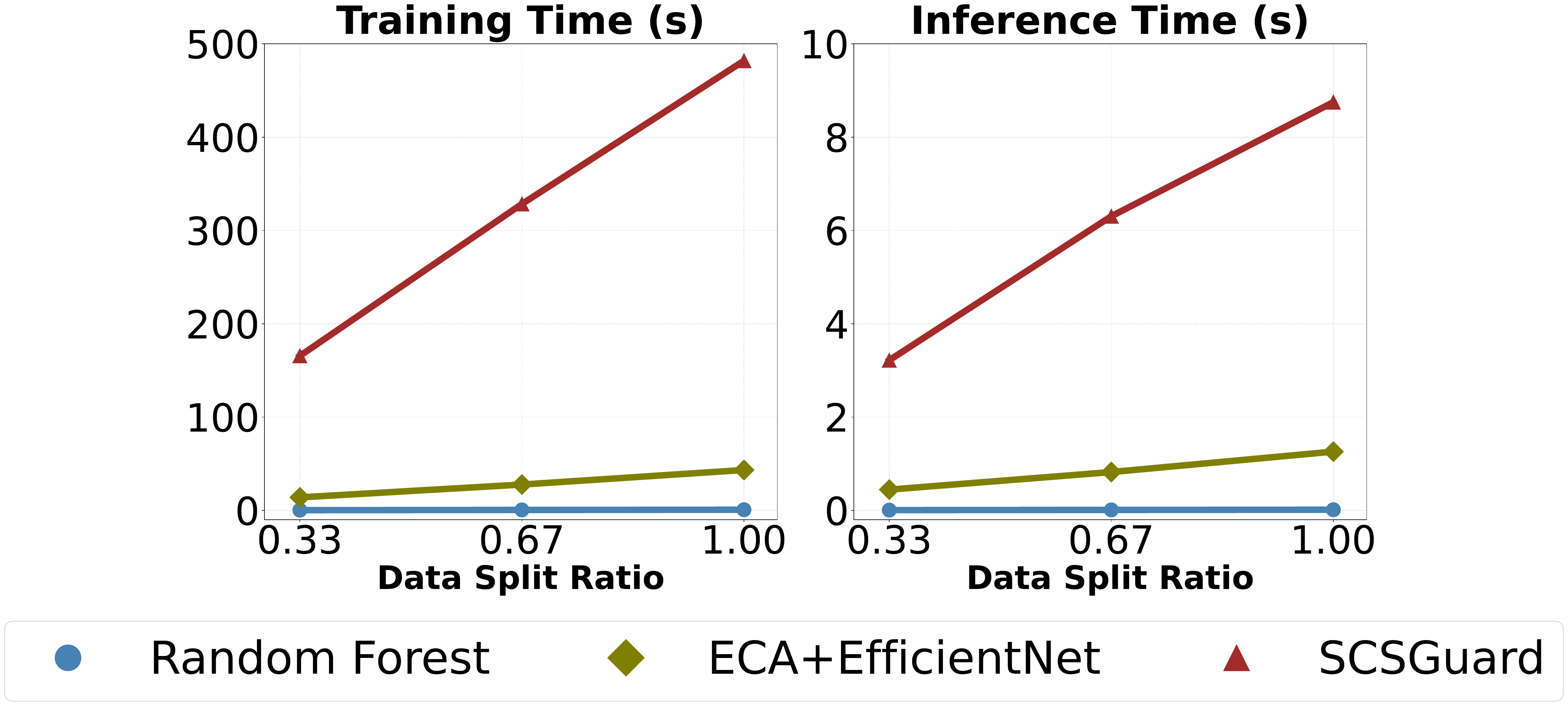}
        \caption{Time metrics of the best models per data split.}
        \label{fig:scaltime}
    \end{minipage}
\end{figure}

Moreover, we adopt the Critical Difference Diagram (CDD)~\cite{DBLP:journals/jmlr/Demsar06} to concisely represent a post hoc analysis on the observed results, shown in \autoref{fig:cdd}.
To produce the CDD, a Friedman test~\cite{friedman} is first conducted to detect whether there are significant differences in the observed model metrics among the data splits.
If some differences are detected, the second step is to perform a Wilcoxon signed-rank test \cite{wilcoxon} to identify which pair of models exhibit a significant difference. 
The rightmost elements in the CDD are the ones that showed the best performances (Random Forest for all the four metrics). 
The leftmost ones are the least performing models (ECA+EfficientNet). 
The thick horizontal line connects the three classifiers that failed the Wilcoxon test, indicating no statistical evidence.
Across all comparisons, $p$ is 0.25 or 0.75, with $p_{adj}=0.75$. Cliff’s $\delta$~\cite{cliff1993dominance} suggests varying effect sizes, with notably negative values for SCSGuard compared to ECA+EfficientNet (-0.778 for Accuracy and F1 Score, -0.333 for Precision, and -1.0 for Recall), reflecting performance declines. 
However, the high $p_{adj}$ values indicate that these differences are not statistically significant. 
This is likely due to the small sample size in the scalability experiment (36 measurements in total), as non-parametric methods require larger datasets for robust statistical power~\cite{Conover1999}.

We further study the training and inference times for such models, shown in \autoref{fig:scaltime}.
The impact of data augmentation on complex models, in particular for Language models, is extremely costly. 
On average (among all data splits), SCSGuard has a training time of 325.302 seconds (+64733.4\% respect to the Random Forest and +1030.6\% respect to ECA+EfficientNet). The average inference time on one batch of data is 6.091 seconds (+57258.5\% respect to the Random Forest and +622.4\% respect to ECA+EfficientNet).
As we can see, for SCSGuard, both training and inference times almost double (+77.35\% on average training time and +68.8\% on average inference time) each time the data sample is enlarged, while for both HSCs and the VM they remain stable and low. 
Therefore, incrementing the number of contract samples in the dataset can increase significantly the performances of complex models, and especially the Language ones.
However, at the same time, the associated loss in time efficiency should be considered.
The importance of timeliness is generally application-specific: in crypto wallets, users interact with smart contracts in real-time, often signing transactions within seconds. 
Any delay in detecting a phishing contract could mean a user already approved a malicious transaction before getting a warning. 
DeFi projects list new tokens frequently, and malicious projects can rug-pull within minutes while security firms analyze threats over time, so delays aren’t as critical.

\subsection{Time-resistance analysis}
\begin{figure}[!t]
    \centering
    \includegraphics[width=\linewidth]{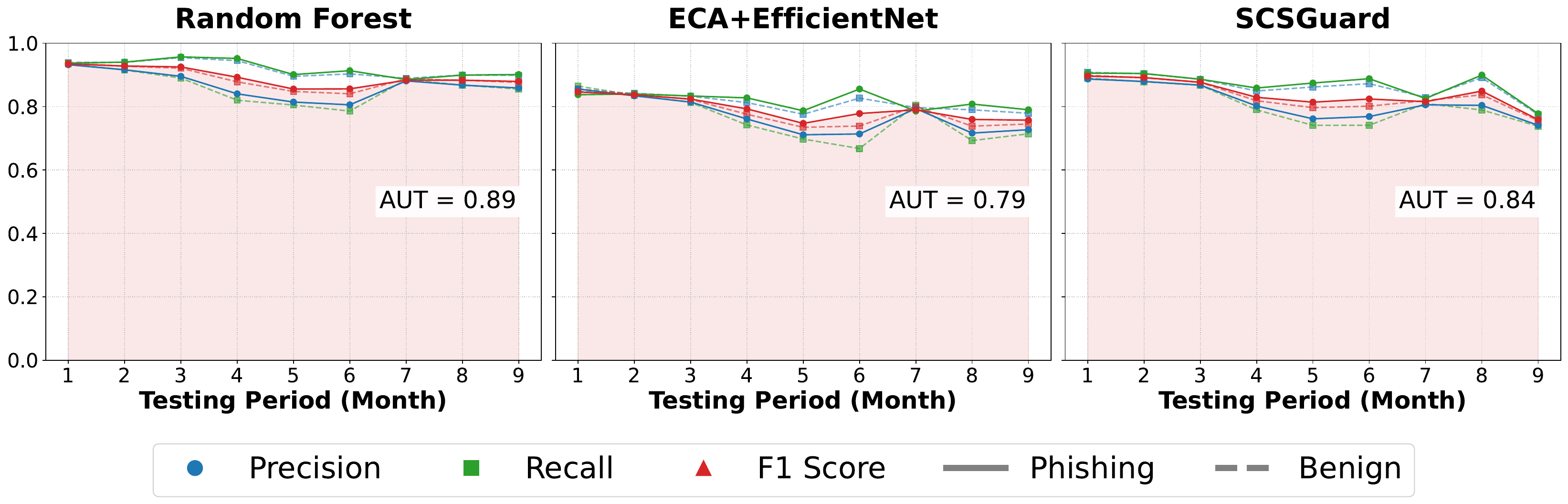}
    \caption{Time evolution of performance metrics over nine months, with AUT for the phishing samples' F1 score.\label{fig:time_resistance}}
\end{figure}

Building on~\cite{DBLP:conf/uss/PendleburyPJKC19}, we design a time-resistance experiment to evaluate \sys's robustness against temporal performance decay. We construct a second dataset of 7,000 samples, ensuring benign samples match the temporal distribution of phishing ones (\autoref{fig:num_contracts}). The training set includes smart contracts from October 2023 to January 2024, while nine test sets, spanning February to October 2024, assess performance over time.
As in the scalability experiment, we evaluate SCSGuard, ECA+EfficientNet, and Random Forest to determine whether phishing contracts exhibit persistent patterns or evolve to evade detection. Results (\autoref{fig:time_resistance}) show stable phishing detection performance, with only a slight decline due to evolving attack patterns, as explained in the reference study.
To quantify stability, we use Area Under Time (AUT), with $\text{AUT} \in [0, 1]$, which represents the area under the F1 curve for phishing samples. A higher AUT indicates greater robustness against evolving threats. Random Forest achieves the highest stability (AUT = 0.89), followed by SCSGuard (0.84) and ECA+EfficientNet (0.79), which experiences more fluctuations. Despite minor variations, \sys effectively adapts to new threats, proving to be a reliable long-term phishing detection solution. 
The dataset for this analysis is publicly available at~\cite{DBLP:software/de-rosa2025}.
\label{sec:eval:time}

\subsection{Influence of opcode prevalence on the best classifier}
\label{sec:eval:shap}

\begin{figure}[!t]
    \centering
    \includegraphics[width=\linewidth]{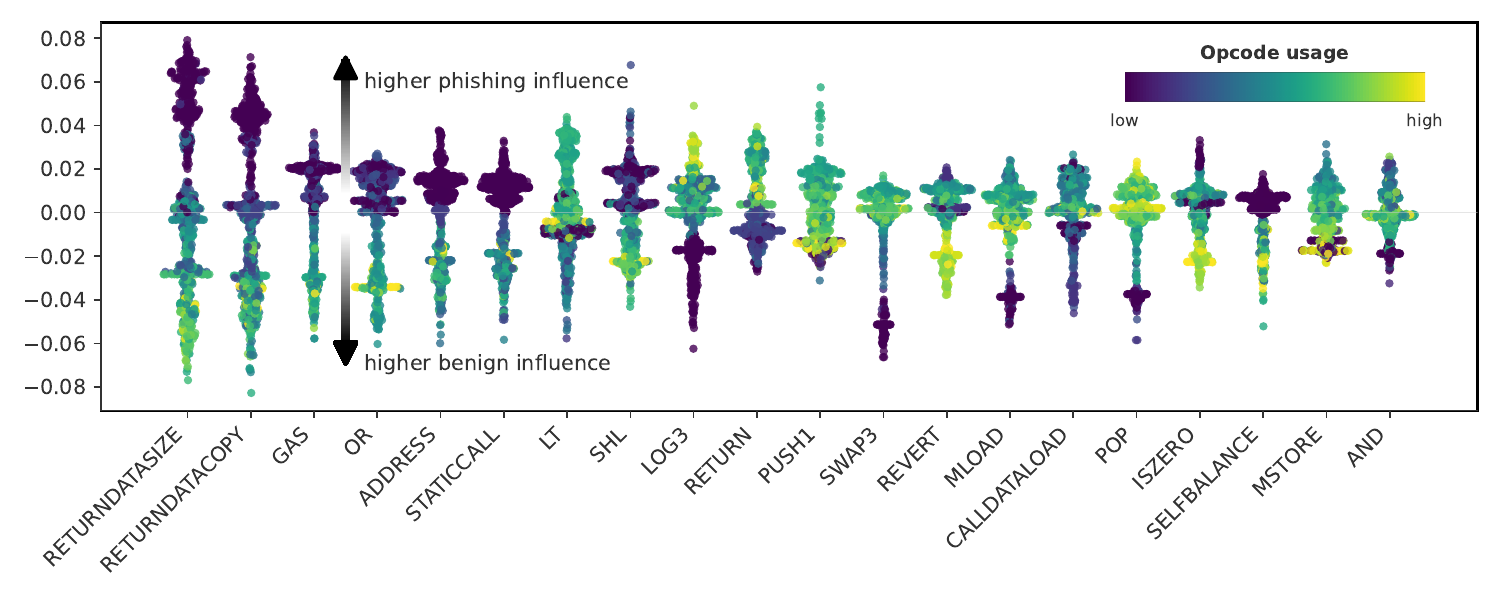}
    \caption{HSC classifier's SHAP values of all samples of a test split (20 most influential features are shown). \label{fig:shap_summary}} 
\end{figure}

Beyond the statistical analysis presented earlier, examining the influence scores of the opcodes provides complementary insights for the model designer. 
Due to space concerns, we restrict ourselves to the best-performing model, HSC with Random Forests. 
\autoref{fig:shap_summary} displays the Shapley values~\cite{DBLP:conf/nips/LundbergL17} associated with the 700 samples that comprise the test set of a random fold from \S\ref{sec:eval:res} (other interpretability tools exist, such as impurity-based feature importance, but they can be biased towards high-cardinality features which can result in misleading importance scores~\cite{scikit-learn-permutation-importance}).
The vertical axis represents the SHAP value, which quantifies the contribution of each feature (here, opcode usage) along the horizontal axis in shifting the model’s prediction for that instance away from the base value (\ie, the mean probability of phishing across all contracts).
For instance, the cluster of contracts with Shapley values of $0.025$ that use GAS very rarely (3$^{\text{rd}}$ column) suggests the classifier finds low usage of GAS suspicious. 
Many well-structured contracts manage gas efficiently, especially when dealing with external calls (\eg, as in {\small{\texttt{Address.functionDelegateCall(address(this),}} \\
\texttt{data[i])}}~\cite{etherscan_0xb5e7b87e7a84276b13da3f07495e18f3e229d3a0})
for which controlled execution may explicitly check the available gas before proceeding.
Phishing contracts, on the other hand, often lack such safety checks because they are designed to steal funds quickly rather than execute complex operations.
However, operations that trigger GAS can be nested inside loops (\eg, a source code can contain 
{\small{\texttt{if (deprecated) \{return UpgradedStandardToken(upgradedAddress).transfer}\\ \texttt{ByLegacy(msg.sender, \_to, \_value);\}}}}~\cite{etherscan_0x279e2f385cE22F88650632D04260382bFB918082}) and still be present in the disassembled bytecode, which can dilute the behavior and render this indicator unreliable (in this case, the contract was a false negative).

%% file: discussion.tex
\section{Discussion and Lessons Learned}\label{sec:lessons}
\label{sec:discussion}
\sys is an efficient framework to systematically train, evaluate and assess ML models to detect phishing activities in the EVM. 
We demonstrate that it is possible to create extremely efficient phishing classification systems ($\approx 90\%$ accuracy across all the models) simply considering the contract bytecode, without collecting transaction traces by interacting with potential malicious nodes in the blockchain.
\begin{tcolorbox}
\small{\textbf{Take-away 1: Leveraging \sys, we can put in place extremely effective opcode-based phishing detection systems.}}
\end{tcolorbox}
To navigate the limited data availability caused by de-duplication of minimal proxy contracts, we developed a post hoc analysis module to assess and generalize results from the $n$ samples collected to the full set $N$ of contracts deployed in the chain. Our findings showed significant differences between models of different categories, while divergences among similar models were less relevant.
\begin{tcolorbox}
\small{\textbf{Take-away 2: LMs and VMs do not perform better than HSCs on phishing detection. Differences in the same categories of models are less meaningful.}}
\end{tcolorbox}
Our scalability analysis highlighted how more complex models overcome simpler HSCs in the long run at the cost of time efficiency. 
This conclusion meets the idea that large models likely require more data to achieve good generalization, which is a staple of machine learning research~\cite{DBLP:journals/corr/abs-1812-11118}.
We assume that there is margin to improve the $\approx 94\%$ accuracy obtained with HSCs by enlarging the dataset. 
\begin{tcolorbox}
\small{\textbf{Take-away 3: Complex models, especially LMs, scale better than HSCs. Increasing the data size, we can even improve the accuracy of the detection system using LMs.}}
\end{tcolorbox}

Finally, we conducted a time-resistance analysis to evaluate \sys's ability to adapt to evolving attack patterns. The results indicate that \sys maintains overall robustness against temporal performance decay. Among the components, HSCs exhibited the highest stability, while LMs and VMs experienced minor fluctuations.
\begin{tcolorbox}
\small{\textbf{Take-away 4: \sys is resilient to evolving threats. Despite minor fluctuations, it effectively adapts to ensure reliable long-term performance.}}
\end{tcolorbox}

%% file: rw.tex
\section{Related work}
\label{sec:rw}
To the best of our knowledge, \sys is the first pure phishing detection framework for smart contracts based on the analysis of the opcodes. 
Some early classification of vulnerable and/or malicious smart contracts exist, such as~\cite{durieux2020empirical}.
We organize related work into four detection system categories: \emph{(A)} opcode-based fraud, \emph{(B)} transaction-based phishing, \emph{(C)} opcode-based vulnerability and \emph{(D)} symbolic execution and verification tools.

\subsection{Opcode-based fraud detection systems}
These systems provide efficient fraud detection systems for smart contracts basing on their bytecode analysis, without specifically address phishing activities. 
Alongside SCSGuard~\cite{DBLP:conf/infocom/HuBX22} and ECA+EfficientNet~\cite{DBLP:journals/cmc/ZhouYWWHL23} (see \S\ref{sec:eval:des}), HoneyBadger~\cite{honeypots} employs symbolic execution and heuristics to discover honeypots in smart contracts. 
It takes as input the EVM bytecode and returns the detected honeypot techniques. 
Similarly, Al-SPSD~\cite{ponzi} leverages ordered boosting on the features extracted from the original opcode sequences to detect Ponzi schemes in smart contracts.
Finally, other approaches~\cite{DBLP:journals/jpscp/LiuPF23, DBLP:journals/access/EhsanSAMADA24} demonstrated the efficacy of supervised learning approaches, such as Random Forests (RF)~\cite{rf} and k-Nearest Neighbors (k-NN)~\cite{knn}.
In both cases, the authors adopt feature extraction techniques on the original opcode sequences, before feeding the data to the classifier.

\subsection{Transaction-based phishing detection systems}
These tools detect phishing activities on smart contracts based on the transaction traces rather than pure bytecode analysis.
Ethereum Phishing Scam Detection (Eth-PSD)\cite{ethpsd} attempts to detect phishing scam-related transactions using several ML classifiers.
TxPhishScope~\cite{txphishscope} dynamically visits suspicious websites, triggers transactions, and simulates results in order to detect phishing websites and extract related accounts automatically. 
Labeling transactions and labeling contracts are fundamentally different tasks. 
While contracts can be indirectly labeled through associated transactions, accurate analysis typically requires many transactions per contract. 
Additionally, replaying transactions tied to malicious actors may expose sensitive user data, creating security and privacy risks.

\subsection{Opcode-based vulnerability detection systems}
The following tools detect smart contract's code vulnerabilities (and not social engineering attacks) based on bytecode analysis, as
ESCORT~\cite{escort} (see \S\ref{sec:eval:des}).
CodeNet~\cite{codenet} is a CNN architecture to classify vulnerable contracts while preserving the semantics and context of the smart contract. 
WIDENNET~\cite{widennet} leverages a variant of the Wide \& Deep neural network architecture~\cite{DBLP:conf/recsys/Cheng0HSCAACCIA16}.
In~\cite{DBLP:journals/access/QianLHZW20}, the authors propose a bidirectional long-short term memory with attention mechanism (BLSTM-ATT), aimed to detect a specific vulnerability, \ie reentrancy bugs.
Finally, in~\cite{bilstm} the authors implement a 2-layer bidirectional LSTM to detect code vulnerabilities in smart contracts.

\subsection{Symbolic execution and verification tools}
Symbolic execution tools can analyze and test smart contracts deployed in the EVM by exploring their possible execution paths to detect specific vulnerabilities and bugs. 
DefectChecker~\cite{DBLP:journals/tse/ChenXLGLC22a} detects eight contract defects that cause unwanted behaviors of smart contracts. 
In~\cite{dia2021empirical}, the authors define a smart contract defect classification scheme to evaluate the effectiveness of three popular verification tools: Mythril~\cite{mythril}, Securify2~\cite{securify2}, and Slither~\cite{slither}. 
The results show that the tools have limited detection effectiveness but are complementary in identifying different types of faults. 
On the other hand,~\cite{almakhour2020verification} provides a comprehensive survey of verification methods, highlighting that most tools handle only simple contracts, with complex ones still posing challenges.
In~\cite{ghaleb2020effective}, the authors propose SolidiFI, a framework to evaluate six popular static analysis tools for smart contracts. 
It identifies numerous undetected bugs despite the tools' claimed capabilities. 
Finally,~\cite{grieco2020echidna} presents Echidna, an open-source smart contract fuzzer designed to efficiently uncover real bugs with minimal user intervention and high execution speed.

%% file: conclusion.tex
\section{Conclusion}
\label{sec:conclusion}
\sys is the first comprehensive framework aimed at detecting phishing attacks on Ethereum smart contracts by analyzing opcodes and bytecode. 
Our extensive evaluation of 16 phishing detection models underscores the efficacy of opcode-based fraud detection, achieving high accuracy and robust performance metrics across a variety of methodologies.
Our results show that more complex models (\ie LMs) do not necessarily yield better results overall but could scale better.
Our key contributions include the development and public release of the largest dataset of phishing smart contracts, a detailed architectural design for the \sys framework, and an in-depth experimental evaluation of multiple detection approaches. 
By providing an open-source, reproducible platform, we aim to foster further research and development in the domain of smart contract security.
Our findings pave the way for future advancements in detecting malicious activities within decentralized environments. 
While the scope of \sys is to detect phishing smart contracts before they are deployed, we consider live detection an interesting future work.
We will explore strategies to efficiently deploy scalable phishing detection systems for smart contracts using \sys. 
Cybersecurity firms or blockchain monitoring platforms, such as Etherscan, are potential customers of our system.